\documentclass[aps,pre,twocolumn]{revtex4}
\usepackage{graphics}
\usepackage{lineno,hyperref}
\usepackage{amsmath}
\usepackage{amssymb}
\usepackage{amsfonts}
\usepackage{graphicx}
\usepackage{color}
\begin{document}
\title{Chimeras in Leaky Integrate-and-Fire Neural Networks: \\
Effects of Reflecting Connectivities}
\author{N. D. Tsigkri-DeSmedt}
\affiliation{Institute of Nanoscience and Nanotechnology, National Center for Scientific Research ``Demokritos'', 15310 Athens, Greece}
\affiliation{Department of Physics, University of Athens, Athens, Greece}
\author{J. Hizanidis}
\affiliation{Institute of Nanoscience and Nanotechnology, National Center for Scientific Research ``Demokritos'', 15310 Athens, Greece}
\affiliation{ Crete Center for Quantum Complexity and Nanotechnology, Department of Physics, University of Crete, 71003 Heraklion, Greece}
\author{E. Sch{\"o}ll }
\affiliation{ Institut f{\"u}r Theoretische Physik, Technische Universit{\"a}t Berlin, Hardenbergstra\ss{}e 36, 10623 Berlin, Germany}
\author {P. H{\"o}vel}
\affiliation{ Institut f{\"u}r Theoretische Physik, Technische Universit{\"a}t Berlin, Hardenbergstra\ss{}e 36, 10623 Berlin, Germany}
\affiliation{ Bernstein Center for Computational Neuroscience Berlin, Humboldt-Universit{\"a}t zu Berlin, Philippstra{\ss}e 13, 10115 Berlin, Germany}
\author{A. Provata \thanks{\emph{Corresponding author:} a.provata@inn.demokritos.gr}}
\affiliation{Institute of Nanoscience and Nanotechnology, National Center for Scientific Research ``Demokritos'', 15310 Athens, Greece} 


%
\date{Received: \today / Revised version: date}
\begin{abstract}{
The effects of nonlocal and reflecting connectivity are investigated in coupled 
Leaky Integrate-and-Fire (LIF) elements, which assimilate the exchange of electrical 
signals between neurons. Earlier investigations have demonstrated that non-local
and hierarchical network connectivity often induces complex synchronization patterns 
and chimera states in systems of coupled oscillators. In the LIF system we show that 
if the elements are non-locally linked 
with positive diffusive coupling in a ring architecture
the system splits into a number of alternating domains.  Half of these domains contain elements, whose
potential stays near the threshold, while they are interrupted by active domains, where the elements 
perform regular LIF oscillations. The active domains move around the ring with constant
velocity, depending on the system parameters. 
The idea of introducing reflecting non-local coupling in LIF networks
originates from signal exchange between neurons residing
in the two hemispheres in the brain. 
We show evidence that this connectivity
 induces novel complex spatial and temporal structures: for relatively extensive ranges
of parameter values the system splits in two coexisting domains, one domain
where all elements stay near-threshold and one where incoherent states develop
with multileveled mean phase velocity distribution.  
}
\end{abstract}
%
%
\maketitle
\section{Introduction}
\label{intro}

 Chimera states,
which are characterized by spatial coexistence of coherent and incoherent oscillators,
were first introduced in 2002 by Kuramoto and Battogtokh \cite{kuramoto:2002} and 
 by Abrams and Strogatz in 2004 \cite{abrams:2004}. These states appear in 
systems of nonlocally coupled oscillators as well as in oscillator networks with complex connectivity. 
In chimera states the coherent parts are characterized by constant mean
phase velocity, while the incoherent ones demonstrate an arc-shaped profile
 of the mean phase velocities
\cite{panaggio:2015,schoell:2016}. A wide range of neuronal models 
have been shown to exhibit chimera states,  
\cite{laing:2001,sakaguchi:2006,omelchenko:2011,hizanidis:2014,omelchenko:2013,omelchenko:2015,hizanidis:2016,andrzejak:2016}
The aim of the current study is to investigate the influence of neuron connectivity patterns on the formation of chimera states. For this reason, we consider two kinds of connectivity, nonlocal and reflecting, in the Leaky Integrate-and-Fire (LIF) model, which is one of the classical models in neuroscience \cite{brunel:2008}.

 The problem of connectivity in networks composed of neuronal oscillators is of fundamental
 interest in neuroscience. The human brain consists of approximately
$10^{10}$ nerve cells (neurons) and each of them is linked with up to $10^4$ other neurons via synapses.
Even with today's most advanced Magnetic Resonance Imaging (MRI) and functional Magnetic Resonance Imaging (fMRI) equipment detailed connectivity patterns are not 
exactly known
 and they are the subject of intense international research, both experimentally and
computationally \cite{poldrack:2015,logothetis:2008}. It is widely believed that high connectivity
is relevant in cognition, intelligence and creativity, while low or interrupted connectivity is
related to mental disorders, such as schizophrenia, Alzheimer or Parkinson \cite{ruiz:2013}.

 Despite the difficulties in mapping neuron connectivity patterns, MRI and fMRI 
studies  have revealed local spatial structures connecting the loci in the human brain 
\cite{hannon:2016,finn:2016}. In particular, in ref. \cite{finn:2016} Finn et al. propose
that it is even possible to identify a person from the ``fingerprints'' of their brain connectivity
patterns, as mapped by fMRI. The experimental output of these studies is often used as
input in numerical experiments aiming to understand the flow of information in the brain.
Information is exchanged between neurons as electrical signals which trigger synchronous, delayed
 or asynchronous responses propagating the information. During these exchanges collective 
neuron response can be understood as a pattern.

One might hypothesize
 that each synchronization pattern describes a ``memory'' (or any cognitive activity)
 and can be recalled and
reproduced later if the same or similar electrical signals arrive. Once the connectivity
is established and remains unchanged in time the same input stimulus will create the same synchronization
 effect in the neurons involved and this effect
corresponds to recalling a ``memory'' pattern. This view explains why ``memories'' fade away
with time: once enough of the synapses break or neurons die (due to aging, or disease, or accident)
 then it is not possible to recreate the synchronization pattern again and thus to ``recall
the memory''. The above view of neuron synchronization patterns in the brain drives the need to explore the diversity of synchronization patterns produced by different connectivity schemes 
in neuron networks. 

The coexistence of coherent with incoherent domains found in chimera states may offer an explanation 
of neurological states such as the unihemispheric sleep of animals
 \cite{rattenborg:2000, rattenborg:2016}. In maze-learning experiments
it has been reported that  rats show
increased neural phase synchronization. Their decisions depend on
previous knowledge of the environment which might have been retrieved 
from memory stores \cite{fell:2011}. Phase syncronisation often coexists with chimera states, especially
in complex neuron connectivity patterns.  

 After many decades of studying locally connected neuron oscillators, with and without time delays, 
recently nonlocal connectivity patterns have caught considerable attention. 
This was driven by the numerous applications of biological or technological interest which 
involve systems (networks) of many interacting components. It was made possible due to the fast development of 
computer resources which allows to tackle reliably networks consisting of a large number of units.
 In the classical nonlocal connectivity
schemes each oscillator is linked to a number of elements $R_L \ge 1$ on the left and $R_R \ge 1$
on the right \cite{kuramoto:2002,abrams:2004,omelchenko:2011}.
Mostly, the symmetric case $R_L=R_R$ was studied but asymmetric coupling ($R_L \ne R_R$) has recently
been used to control the undesired erratic lateral motion of chimera states \cite{omelchenko:2016}.
The case $R_L=R_R=1$ corresponds to the local, nearest neighbor connectivity.
The extension of connectivity to regions $R_L,R_R >1$  led to the interesting phenomenon of
chimera and in 2013 to multichimera states\cite{omelchenko:2013}. 
The interplay of
nonlocal connectivity with the nonlinearity of the dynamical scheme has led to multiple incoherent domains
and chimera death states \cite{zakharova:2014,schneider:2015,banerjee:2015,loos:2016,banerjee:2016}.
In the recent years
 connectivity was further
complexified to take into account results of diffusion tensor MRI analysis indicating that the 
neuron axons distribution in the human brain (and thus the neuron connectivity) has hierarchical
(fractal) connectivity \cite{katsaloulis:2009,expert:2011,katsaloulis:2012}. These studies have revealed
the pattern of nested chimeras \cite{omelchenko:2015} which also appears in systems of
hierarchically arranged reactive units
\cite{hizanidis:2015} and coupled Van der Pol oscillators \cite{ulonska:2016}. Other nonlocal schemes
include 2D (square lattice) and 3D (cubic lattice)  connectivity with corresponding chimera patterns 
and are recently reported in \cite{omelchenko:2012b,maistrenko:2015}.

 In previous studies synchronization phenomena have been explored 
in LIF networks using negative (repulsive) diffusive 
coupling with and without refractory period \cite{tsigkri:2015,tsigkri:2016}. 
We showed that in classical nonlocal coupling multichimera states emerge 
whose multiplicity (number of coherent/incoherent domains)
 depends both on the coupling strength and on the refractory period. 
We also showed that hierarchical topology in the coupling induces nested chimera states 
and transitions between multichimera states with different multiplicities. 
These results revealed new complex patterns and dynamical transitions between different multichimera states resulting from the combined effects of nonlinear dynamics with the hierarchical coupling.
In the current study we use a different coupling pattern, the reflecting connectivity, in which each neuron
is connected with a number of other neurons across a standard mirror diagonal of the ring. This idea
stems from studies in the brain where neurons located on the surface of one hemisphere are connected
with neurons arranged on the opposite hemisphere \cite{finn:2016}. We give numerical evidence that
such arrangements lead to multileveled incoherent states
 which coexist with near-threshold elements.
Multileveled incoherent states are characterized by groups of elements having different mean phase 
velocities.
These are novel interesting patterns and occur in LIF neurons when connected via reflecting coupling.

 This work is organized as follows: 
In section \ref{sec:LIF} we recapitulate the main  properties of  
the LIF model, and we study synchronization phenomena in coupled LIF elements with diffusive coupling
when a finite refractory period is introduced. In section \ref{sec:nonlocal-coupling} 
we scan the parameter space and study the formation of
domains with near-threshold elements and clustering phenomena. In section \ref{sec:reflecting-coupling} 
we investigate the effects of introducing reflecting connectivity matrices and 
study the resulting oscillation patterns and configuration transition phenomena. 
We give evidence of transitions between synchronous oscillations and
chimera states with multileveled mean phase velocities,
using as control parameters the coupling strength and the
refractory period. 
In the concluding section we discuss possible applications of this study,
we summarize our main results and discuss open problems.

\section{Leaky Integrate-and-Fire Model with diffusive coupling and finite refractory period}
\label{sec:LIF}

The single neuron LIF model is common for excitatory and inhibitory dynamics and
consists of a single linear differential equation describing the integration of the neuron
potential $u(t)$, followed by a highly nonlinear, abrupt resetting of the potential 
to its ground value $u(t)=u_0$ \cite{tsigkri:2015}. Namely,
\begin{subequations}
\begin{equation}
\frac{du(t)}{dt}=\mu-u(t)  
\label{eq1a} 
\end{equation}
\begin{equation}
\lim_{\epsilon \to 0}u(t+\epsilon ) \to u_0, \>\>\> {\rm when} \>\> u(t) \ge u_{\rm th}.
\label{eq1b}
\end{equation}
\label{eq01}
\end{subequations}
\noindent 
Equation~(\ref{eq1a})
 represents the integration of the potential $u(t)$, which develops accumulating contributions with rate $\mu$ \cite{ermentrout:1998,vilela:2009,lindner:2002,kouvaris:2010}. The term $-u(t)$
on the right hand side of this equation represents the ``leaky term'' which prohibits the 
neuron potential from increasing to arbitrarily large values, but forces it to converge to the
value $\mu$, for $t \to \infty$. An integral constituent of the single LIF model is Eq.~(\ref{eq1b}),
which restores the potential $u(t)$ to its ground value $u_0$ every time the potential exceeds
 its upper threshold
termed as ``$u_{\rm th}$''. This way a closed potential loop is created with extreme values
$u_0 \le u(t) \le u_{\rm th}$.
Due to the structure of Eq.~(\ref{eq1a}), $u(t)$ can reach the value  $u_{\rm th}$ 
and thus drop to the ground state $u_0$ only if $u_{\rm th} < \mu$.

Equation~(\ref{eq1a}) is linear and can be solved analytically as long as the potential
$u(t) \le u_{\rm th}$.  The solution is:

\begin{equation}
u(t)=\mu -(\mu-u_0)\exp {(-t)}, \>\>\>\> {\rm for} \>\>\>  u(t)\le u_{\rm th},
\label{eq02}
\end{equation}
\noindent where, without loss of generality, we have assumed that at $t=0$ 
the potential is at the ground state
 $u_0$.  The  period $T_s$ of the single neuron potential loop
 takes the following form, provided that the resetting to the ground potential is abrupt (has
 zero duration):
\begin{equation}
T_s=\ln \frac{\mu -u_0}{\mu - u_{\rm th}}.
\label{eq03}
\end{equation}
\noindent This period accounts exactly for the duration for the potential of a single (uncoupled) neuron 
to reach 
the threshold value $u_{\rm th}$ starting from the ground state $u_0$.
 Because $ u_0 < u_{\rm th} < \mu $, the period $T_s$ is always positive. Taking the constraints into
account the general solution of the single LIF neuron can be written as:
\begin{eqnarray}
\nonumber
u(t) &=& \left[ \mu -(\mu-u_0)\exp {(-t_{T_s}) }\right] \left[ 1-\delta (t-nT_s)\right] , \\
t_{T_s} &=& t \mod T_s \\
n &=& \frac{t-t_{T_s}}{T_s}.
\nonumber
\label{eq031}
\end{eqnarray}

  Biological neurons spend a resting period after firing each spike, which is called  ``refractory period'', $p_r$. This is an essential feature of spiking neurons and we account for this by adding to the
model Eq.~(\ref{eq01}) the following condition \cite{tsigkri:2015}:

\begin{eqnarray}
\nonumber
u(t)\equiv u_0,  \> \forall t &:& [l (T_s+p_r)+T_s] \le t \le [(l+1)(T_s+p_r)], \\
& & {\rm where} \>\> l=0,1,2,\cdots
\label{eq04}
\end{eqnarray}
\noindent I.e. after each firing at times $lT_s$ the neuron retains its potential to the resting state
 $u_0$ for an additional time interval $p_r$.
With this additional condition, the total period $T$ of the single neuron 
and corresponding frequency $\omega$, including refractory time, are:
\begin{subequations}
\begin{equation}
T=T_s+p_r=\ln \frac{\mu -u_0}{\mu - u_{\rm th}}+p_r,
\label{eq5a} 
\end{equation}
\begin{equation}
\omega =\frac{2\pi}{T}=\frac{2\pi}{\ln \frac{\mu -u_0}{\mu - u_{\rm th}}+p_r}.
\label{eq5b}
\end{equation}
\label{eq05}
\end{subequations}

 Having defined the single neuron dynamics we proceed by placing single neurons in a ring network
and coupling them via two different schemes. The coupling topology as well as the coupling form are
very important in neuron networks. Hereafter, we will use diffusive coupling, which is the simplest
linear coupling and takes the form $[u_j(t)-u_i(t)]$ between neurons $j$ and $i$. The
coupling topology of the brain is very complex, as was
discussed in the Introduction, hence in this study we will use two types of representative coupling:
a) nonlocal coupling in section \ref{sec:nonlocal-coupling} and b) reflecting coupling in section
\ref{sec:reflecting-coupling}. The topology is defined by the adjacency or
connectivity matrix $\sigma(i,j)$  which
is different in cases a) and b) as will be discussed in the corresponding sections.
For a number of $N$ LIF elements, each of them having potential $u_i(t)$, $i=1, \cdots N$ and
connected in a ring topology the network dynamics is described by the following scheme:
\begin{subequations}
\begin{equation}
\label{eq6a} 
\frac{du_i(t)}{dt}=\mu-u_i(t)+\frac{1}{N_i} \sum_{j=1}^{N_i} \sigma (i,j)
\left[ u_j(t)-u_i(t)\right] 
\end{equation}
\begin{equation}
\lim_{\epsilon \to 0}u_i(t+\epsilon ) \to u_0, \>\>\> {\rm when} \>\> u_i(t) \ge u_{\rm th}
\label{eq6b}
\end{equation}
\label{eq06}
\end{subequations}
\noindent where $N_i$ is the total number of elements that are linked to element $i$. In general,
the number of elements $N_i$ to which element $i$ is linked may be different. In Eq.~(\ref{eq06}) 
we assume that all elements have as common parameters (properties) the refractory
period $p_r$, the threshold potential $u_{\rm th}$, the accumulating rate $\mu$, while they start from
different (usually random) initial states. Also in the current work
we assume that the number of linking
sites $N_i=N_c$ is kept constant for all elements. In later studies, these conditions need to be relaxed
since in realistic biological neurons each neuron has different characteristics corresponding to
its role in the brain functions \cite{loos:2016,omelchenko:2015}.
 The $+$sign in front of the coupling term in Eq.~(\ref{eq6a}) defines the attracting coupling
dynamics. When $u_i$ takes high values (higher than $u_j$) the coupling  term  
becomes negative. This contribution tends to decrease the value of $u_i$ (towards $u_j$)
inhibiting its further growth. In the opposite case, when  $u_i$ takes lower values than $u_j$
 the coupling  term  $u_j(t)-u_i(t)$
becomes positive and this contribution tends to increase the value of $u_i$, towards $u_j$. 
Because the coupling term tends to bridge the differences between the interacting
units we regard
this coupling as ``attracting''. 

 Synchronisation properties are quantitatively described by the mean phase velocity $\omega_i$
 of element $i$. If the system is integrated for time $\Delta T$, we compute the number of full
cycles $c_i$
which element $i$ has completed during $\Delta T$.
 The mean phase velocity of element $i$ is then estimated as \cite{omelchenko:2013}: 
\begin{eqnarray}
\omega_i=\frac{2 \> \pi \> c_i}{\Delta T}
\label{eq07}
\end{eqnarray}
\noindent Visually, synchronization patterns and chimera states are represented by the space-time plots.
Space-time plots are colour coded plots that show the evolution of the potential of the oscillators in time and in space. These are particularly useful in the cases of transitions between different
synchronization patterns, when the mean phase velocities do not give meaningful results.

 Specifically for the LIF model it is necessary to introduce an additional quantitative index $A$ which  
represents the ratio of elements below-threshold. This will become clearer in section
\ref{sec:nonlocal-coupling} and Figs. \ref{fig-nl1}, \ref{fig-nl5} and \ref{fig-nl3}  where
 static and oscillatory elements coexist in the system. 
These partially 
static profiles indicate that many elements are confined near-threshold, while a
few escape toward the resting potential $u_0=0$, see for example Fig.~\ref{fig-nl1} (left panels).  
However, spacetime plots indicate that these
``active'' elements are not localized but the activity propagates from one element to the next
 around the ring. Thus the mean-phase velocity is not an appropriate  measure of coherence, because each
element spends some intervals of time near-threshold and other intervals in the ``active'' state 
(excursions below-threshold). Because in these cases the average number of below-threshold elements
is statistically constant in time the representative quantitative index $A$ of elements
below-threshold is defined as:
\begin{eqnarray}
A=\frac{1}{N \Delta T }\sum_{t=1}^{\Delta T }\sum_{i=1}^N q_i(t) 
\label{eq08}
\end{eqnarray}
\noindent where
\begin{eqnarray}
q_i(t)=  
\left\{
  \begin{aligned}
    1 &\>\> {\rm when} \>\> u_i(t)< u_{\rm th}  \\
    0 &\>\> {\rm otherwise}.
  \end{aligned}
\right.
\label{eq09}
\end{eqnarray} 
The index $A$ counts the average number of elements that stay below-threshold and is
calculated after the system has reached  the steady state. Because it counts the number of
active elements it is also 
referred to as ``activity factor''.
 Synchronisation properties in LIF coupled elements were studied in 2010 by Lucioli et al.
 \cite{lucioli:2010} and Olmi et al. \cite{olmi:2010}. 
These studies have demonstrated the presence of chimera states in non-locally 
coupled LIF elements with time delays. In our work we further introduce
a finite refractory period which is a well established characteristic of the biological neuron, and
we also complexify the connectivity, which introduces the novel effect of the
multileveled incoherent domains, as we will see in the section~\ref{sec:reflecting-coupling}. 

 Without loss of generality, throughout this work the working parameters
are kept to values $\mu =1.0$, $u_{\rm th}=0.98$, $u_0=u_{\rm rest}=0$. 
The system size is kept to $N=1000$ elements.

\section{Effects of nonlocal diffusive coupling}
\label{sec:nonlocal-coupling}

 In the case of nonlocal connectivity of the  LIF elements
arranged in a ring geometry,
the matrix $\sigma (i,j)$ takes the form:
\begin{eqnarray}
\sigma (i,j)=  
\left\{
  \begin{aligned}
    \sigma &\>\> {\rm for} \>\> i-R<j<i+R  \\
    0 &\>\> {\rm otherwise}
  \end{aligned}
\right.
\label{eq10}
\end{eqnarray}
\noindent where $\sigma$ is positive and 
all indices are understood $\mod (N)$.  In this case the element $i$ is linked with $R$ other 
elements to its left and $R$ elements to its right. The 
number of connections $N_i=N_c=2R$, is equal for all elements.
The dimensionless coupling range is defined as $r=R/N$ and
 represents the fraction of the system's constituents
contributing to the dynamics of each element.
 In the next subsection we present the synchronization phenomena
observed as we vary the system parameters, the coupling range $r$, the
coupling strength $\sigma$ and the refractory period, $p_r$.

\subsection{Variation of the coupling range}

Using the nonlocal connectivity rules of Eqs.~(\ref{eq06}) we simulate a LIF ring network
of $N=1000$ oscillators and record the evolution of profiles and spacetime plots.

\begin{figure}[h!]
\includegraphics[clip,width=0.9\linewidth,angle=0]{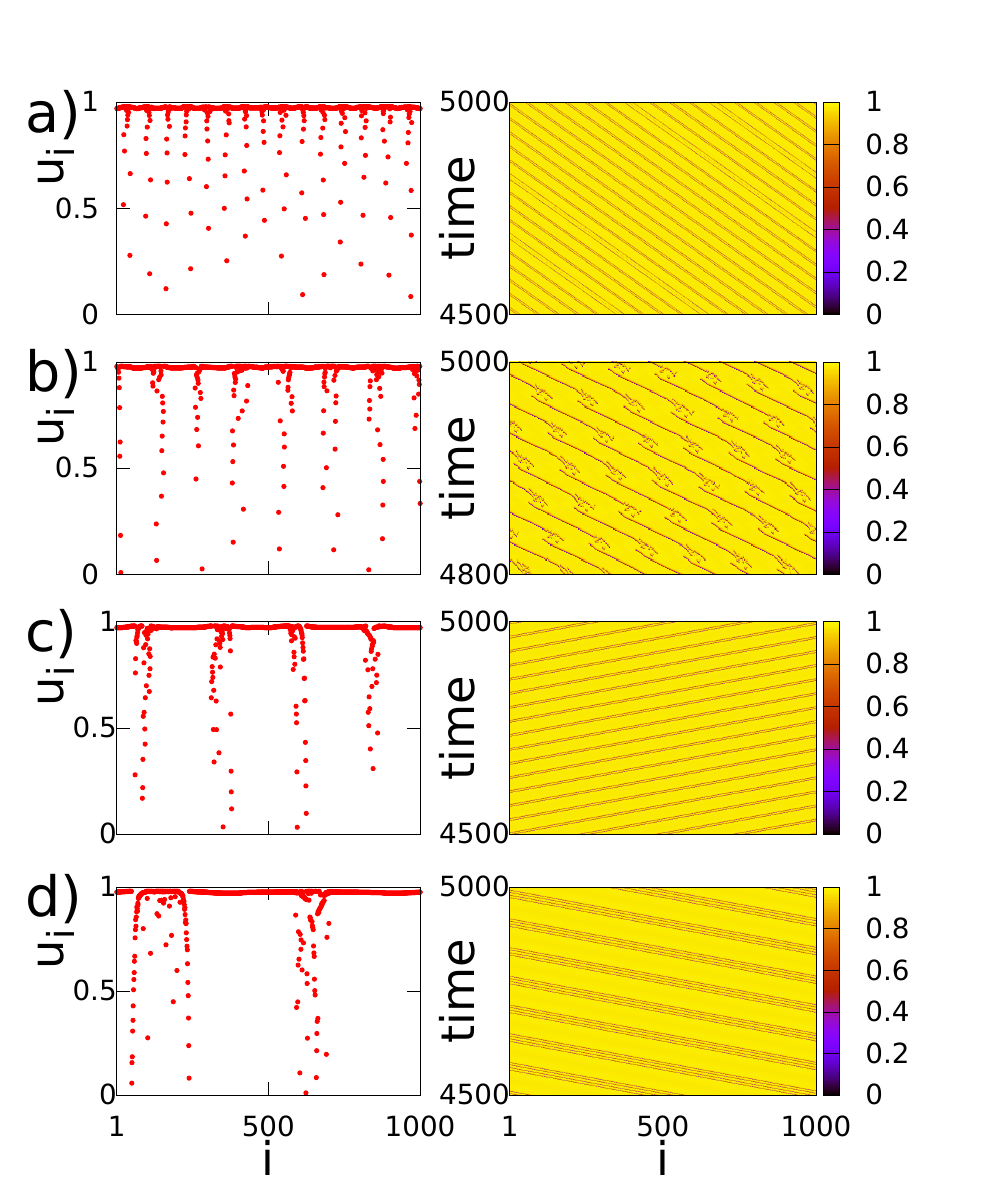}
\caption{\label{fig-nl1} (Color online)
LIF system with nonlocal  connectivity: Typical snapshots (left column)
and spacetime plots (right column).   a) $R=50$ , b) $R=100$, c) $R=200$, d) $R=350$.
Other parameters are  $p_r=0$,  $\sigma =0.7$,   $N=1000$,  $\mu =1.$ and $u_{\rm th}=0.98$.
All realizations start from the same initial conditions, randomly chosen between 0
and $u_{\rm th}$.
}
\end{figure}
\begin{figure}[!h]
\includegraphics[clip,width=0.85\linewidth,angle=0]{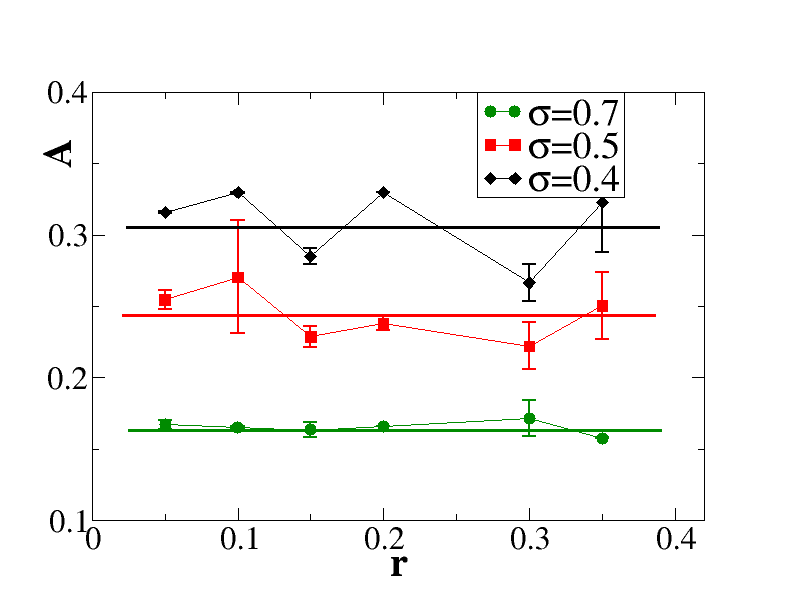}
\caption{\label{fig-nl2} (Color online)
The activity factor $A$ in the 
LIF system with diffusive nonlocal  coupling as function of the coupling range $r=R/N$
and for $\sigma =0.4$ (black line)  $\sigma =0.5$ (red line) and $\sigma =0.7$ (green
line). Averages are taken over 10 initial conditions. The straight lines are guides to the eye.
Other parameters as in Fig.~\ref{fig-nl1}.
}
\end{figure}

 Figure~\ref{fig-nl1}  presents typical snapshots and spacetime plots of the system for different
values of the coupling radius $R$ and keeping the other parameters constant. 
All realizations shown in Fig.~\ref{fig-nl1} 
 start from the same initial conditions, $u_i(t=0), \>\>\> i=1,\cdots N$,
 randomly chosen between 0 and $u_{\rm th}$.
The snapshots and spacetime plots show that the oscillators organize into spatial clusters
where all elements remain close to the threshold while these regions are interrupted 
by oscillating regions
in which the elements make excursions away from the threshold. In most cases
examined the oscillating regions travel through the system with constant velocity. As the moving
fronts on the right panels indicate, the velocity of the
oscillating regions around the ring changes inversely with the parameter $r$.

 Figure \ref{fig-nl1}a shows the formation of active domains alternating with domains where the
elements stay near-threshold. The active (and near-threshold) domains move around the 
ring with constant velocity, as is indicated by the constant slopes in the corresponding (right)
spacetime plot.
As  the coupling range $r$
increases, Fig.~\ref{fig-nl1}b,c, nearby oscillating domains merge into larger active clusters which are
interrupted by threshold elements staying near potential $u_{\rm th}$. This merging 
reduces the number of active $ N_{\rm a} $ 
and near-threshold $ N_{\rm th} $ domains, without violating  
(statistical) continuity around the ring. If we denote by $\langle N_{\rm a}\rangle$ the average number of
active domains and $\langle s_{\rm a}\rangle$ the average size of each active domain, and similarly for the 
domains containing threshold elements, then the continuity condition around the ring is 
expressed by the conditions: 
\begin{equation}
\begin{split}
N_{\rm th}=N_{\rm a} \\
\langle N_{\rm th}\rangle \langle s_{\rm th}\rangle+\langle N_{\rm a}\rangle \langle s_{\rm a}\rangle=N
\label{eq11}
\end{split}
\end{equation}
where $s_{\rm th}$ and $s_{\rm a}$ are the sizes of the near-threshold and active domains,
respectively. The size of the active clusters
as well as the size of the near-threshold regions stay (statistically) constant in time. 
The  
oscillatory motion observed in Figs. \ref{fig-nl1}a,c,d, i.e. 
a number of waves travelling at constant velocity in a sea (1D ring) of almost immobile elements, 
is reminiscent of 
soliton propagation in a medium, although the condition for solitons are not exactly met
in this system. The difference with soliton phenomena is apparent in Fig.~\ref{fig-nl1}b,
where the moving wave disappears while a new one is created in a nearby node. These patterns
correspond to  intermediate states which appear when the number of active (and near-threshold)
domains changes and will be discussed further in section \ref{sec:nonlocal-coupling}B.


 A counter-intuitive observation is that the number of active (below threshold) elements does not significantly  
depend on $r$. This can be seen in Fig.~\ref{fig-nl2} where the fraction of below-threshold
elements is calculated as a function of $r$. As a result, (see Fig.~\ref{fig-nl1}b,c,d)
 as $r$ increases
the elements are redistributed into equidistant clusters, 
while the distance (in threshold elements) between active
clusters increases.
 More specifically, in Fig.~\ref{fig-nl2} we present the fraction $A$ of elements 
which are below the threshold as a function of $r$, for three different values of the
coupling constant $\sigma$. The value $A$ is calculated as the average (over time)
fraction of elements which have potential $u_i \le 0.97$. Average values (and error bars)
are taken over 10 different initial conditions. In cases where the error bars are not 
visible, their size is smaller that the representing symbol. 
We observe that $A$ stays constant, up to statistics. This means
that the magnitude of activity in the system mostly depends on the strength of the coupling 
constant. While the coupling strength $\sigma$ is related to the magnitude of the
activity, the coupling range $r$ is related to the organization of the activity in isolated
regions and this might have important implications in the exchange of electrical 
signals between neurons. 

\subsection{Variation of the coupling strength}

\begin{figure}[ht!]
\includegraphics[clip,width=0.9\linewidth, angle=0]{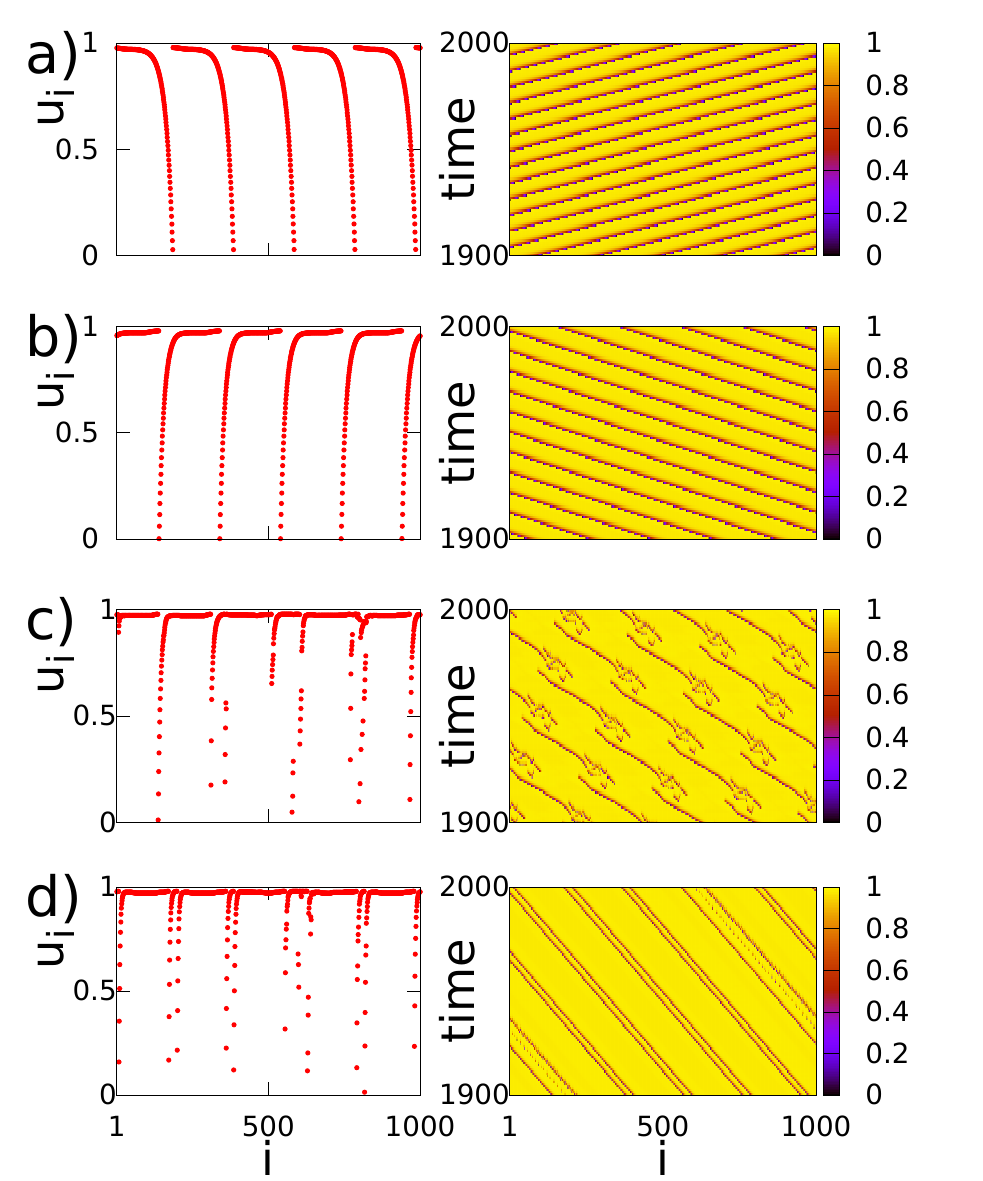}
\caption{\label{fig-nl5} (Color online)
LIF system with nonlocal  connectivity: Typical snapshots (left column)
and spacetime plots (right column).   a) $\sigma =0.2$ , b) $\sigma =0.3$ , 
c) $\sigma =0.6$ and d) $\sigma =0.7$.
Other parameters are  $p_r=0$,  $R=150$,  $N=1000$,  $\mu =1.0$ and $u_{\rm th}=0.98$
}
\end{figure}

\begin{figure}[h]
\includegraphics[clip,width=0.85\linewidth, angle=0]{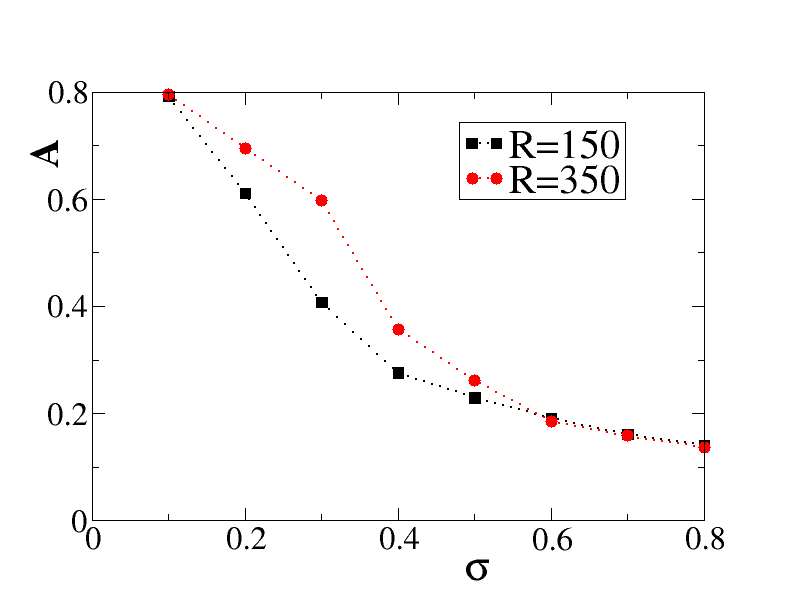}
\caption{\label{fig-nl6} (Color online)
The activity factor $A$ in the 
LIF system with diffusive nonlocal  coupling as function of the coupling constant $\sigma$,
for two different values of the coupling radius $R=150$ (black squares)  and $R=350$ (red circles).
Other parameters as in Fig.~\ref{fig-nl1}.
}
\end{figure}

   Modifications to the general activity of the system is also achieved with variations in the
 coupling constant $\sigma$.
In Fig.~\ref{fig-nl5}a, for small values of $\sigma$ we observe synchronized waves propagating
through the system. 
 As the coupling strength increases the oscillators organize into near-threshold clusters. These clusters are interrupted
by elements making 
abrupt excursions towards the resting potential $u_0=0$ (Fig.~\ref{fig-nl5}b).  While for small
values of $\sigma$ the oscillators follow one-another keeping a constant phase difference (Fig.~\ref{fig-nl5}
a,b), for large values of the coupling constant clusters are formed which travel around the 
ring with a constant velocity (Fig.~\ref{fig-nl5}d). The transition from the synchronized waves to
the clusters occurs for intermediate parameter values, around $\sigma \sim 0.6$ for
this working parameter set, as  depicted in Fig.~\ref{fig-nl5}c.
Here a wave disappears, but before disappearing it coexists with another one which has
spontaneously been created at  a nearby location. These hybrid states mediate (are at the turning
point) between the simple moving waves and the clustered ones. As the parameter $\sigma$
increases above $0.6$ the regions of coexistence increase, forming gradually the clusters.
These travelling clusters are separated again by near-threshold domains. The same  effect
was previously seen in Fig.~\ref{fig-nl1}b, between the  simple active  clusters (Fig.~\ref{fig-nl1}a)
and the composite (grouped) ones (Fig.~\ref{fig-nl1}c and d). For the working parameters used in 
Fig.~\ref{fig-nl1} the transition was recorded around
$R=100, \>\> \sigma =0.7, \>\> p_r=0$.

 The activity factor $A$ as a function of $\sigma$
is shown in Fig.~\ref{fig-nl6}. We note a decrease in the number of elements making excursions
below the threshold as the coupling strength grows. This can be intuitively understood as follows:
The tendency of the oscillators to stay near the threshold is attributed to the exponential
nature of the dynamics which tends asymptotically to $\mu$ as $t\to\infty$. 
As $\sigma$ increases  the system has the tendency to
synchronize and thus more and more elements are attracted to stay near the threshold,
and thus $A$ decreases accordingly.
 As seen from Fig.~\ref{fig-nl6} confirming Fig.~\ref{fig-nl2},
 the value of the coupling range $r=R/N$ does not play a
significant difference in the scaling of $A$.
 We note here that the splitting of the system in near-threshold
and active domains is only observed in the case of attracting coupling
where the elements are held to stay together near the threshold. This
phenomenon is not observed in repulsive coupling, where more classic
chimeras are supported \cite{tsigkri:2015,tsigkri:2016}.

\subsection{Variation of the refractory period}

\begin{figure}[ht!]
\includegraphics[clip,width=0.9\linewidth,angle=0]{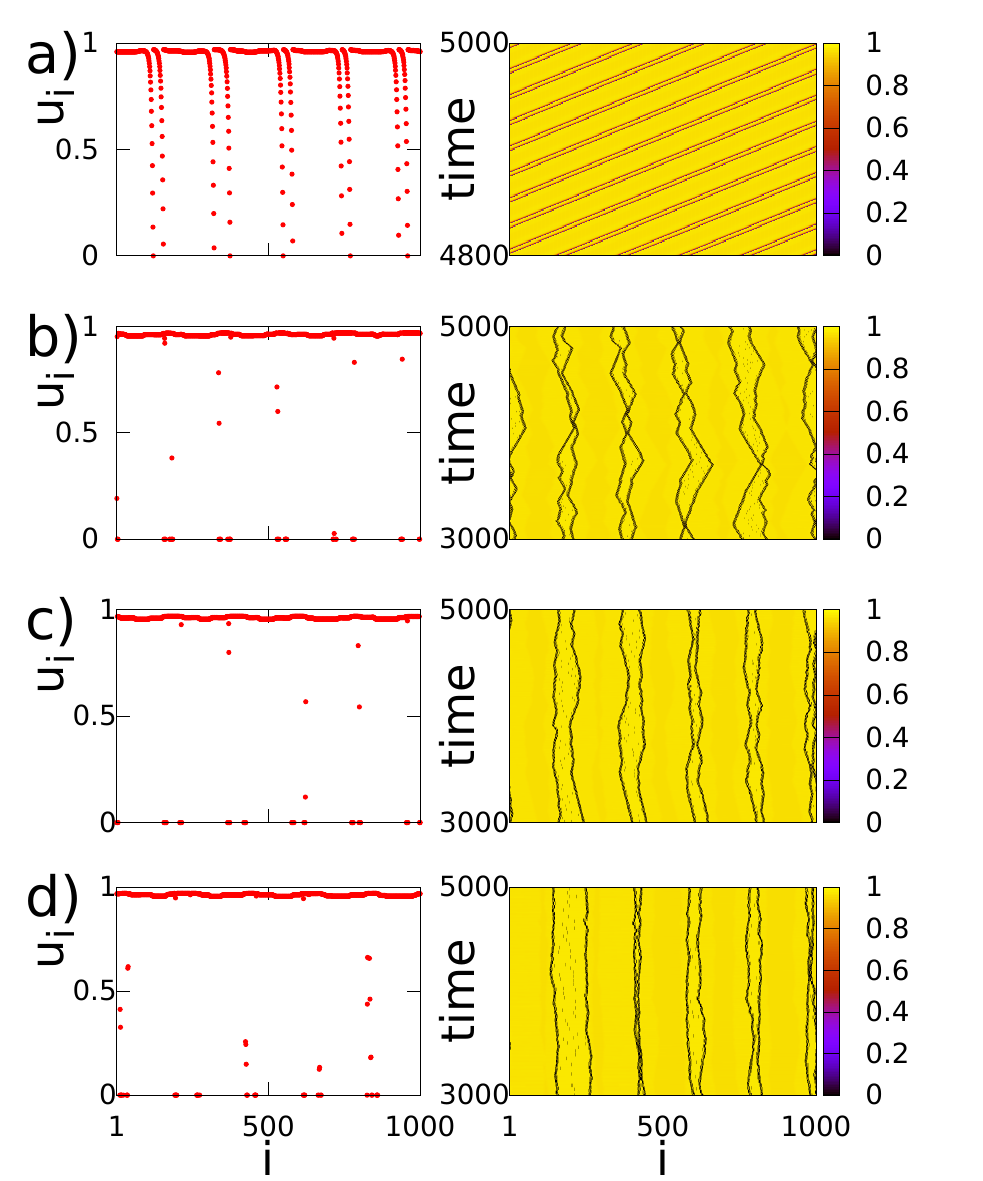}
\caption{\label{fig-nl3} (Color online)
LIF system with nonlocal  connectivity: Typical snapshots (left column)
and spacetime plots (right column).   a) $p_r=0.01 Τ_s$, b) $p_r=0.3 T_s$, c) $p_r=0.5 T_s$ 
and d) $p_r=0.8 T_s$.
Other parameters are  $R=150$,  $\sigma =0.7$   $N=1000$,  $\mu =1.0$ and $u_{\rm th}=0.98$.
}
\end{figure}

 Similarly to Figs. \ref{fig-nl1} and Fig.~\ref{fig-nl5}, Fig.~\ref{fig-nl3}
displays variations in the snapshots and spacetime plots for different values of the refractory
period, keeping all other parameters fixed. For these plots a constant $R=150$ value was used.

\begin{figure}[!h]
\includegraphics[clip,width=0.85\linewidth, angle=0]{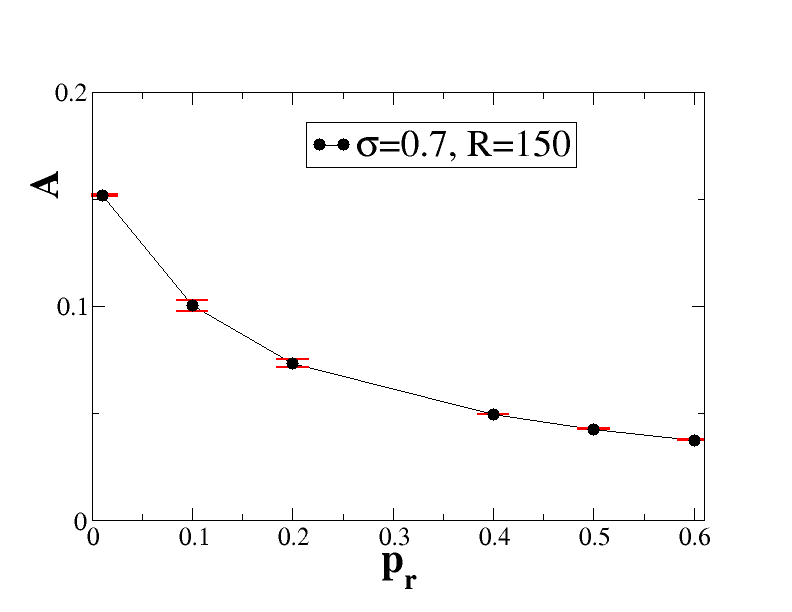}
\caption{\label{fig-nl4} (Color online)
The activity factor $A$ in the 
LIF system with diffusive nonlocal  coupling as function of the refractory period $p_r$,
for coupling radius $R=150$. Averages are taken over 10 initial conditions.
Other parameters as in Fig.~\ref{fig-nl3}.
}
\end{figure}

The opposite effect is seen as the refractory period increases away 
from 0.  For small
values of $p_r$ and for constant values of sigma $\sigma =0.7$,  Fig.~\ref{fig-nl3}a shows clustering of 
active oscillators. As also discussed in previous sections, \ref{sec:nonlocal-coupling}A and B,
the clusters travel around the ring with a constant velocity as is clearly indicated by the
corresponding spacetime plot. At the same time, all other elements take near-threshold values.
As $p_r$ increases the
 ratio of below-threshold elements $A$ decreases, while the position of the clusters gets pinned in space, 
see Fig.~\ref{fig-nl3}b,c. The activity is anchored around specific points of the ring,
 fluctuating locally around them. As $p_r$ increases the fluctuations decrease and as can be seen in 
 Fig.~\ref{fig-nl3}d,  only specific elements make excursions away from the threshold.
The decrease in the system activity is shown in Fig.
\ref{fig-nl4}, which displays the change of $A$ as a function of $p_r$. Averages are
taken over 10 different initial conditions. The 
refractory period induces a gradual, small decrease in the activity of the system,
while pinning the position of the active elements. This effect
has been verified for other values of $r$ and $\sigma$ (not shown).

\section{Effects of reflecting coupling}
\label{sec:reflecting-coupling}

\begin{figure}[ht!]
\center{
\includegraphics[clip,width=0.6\linewidth,angle=0]{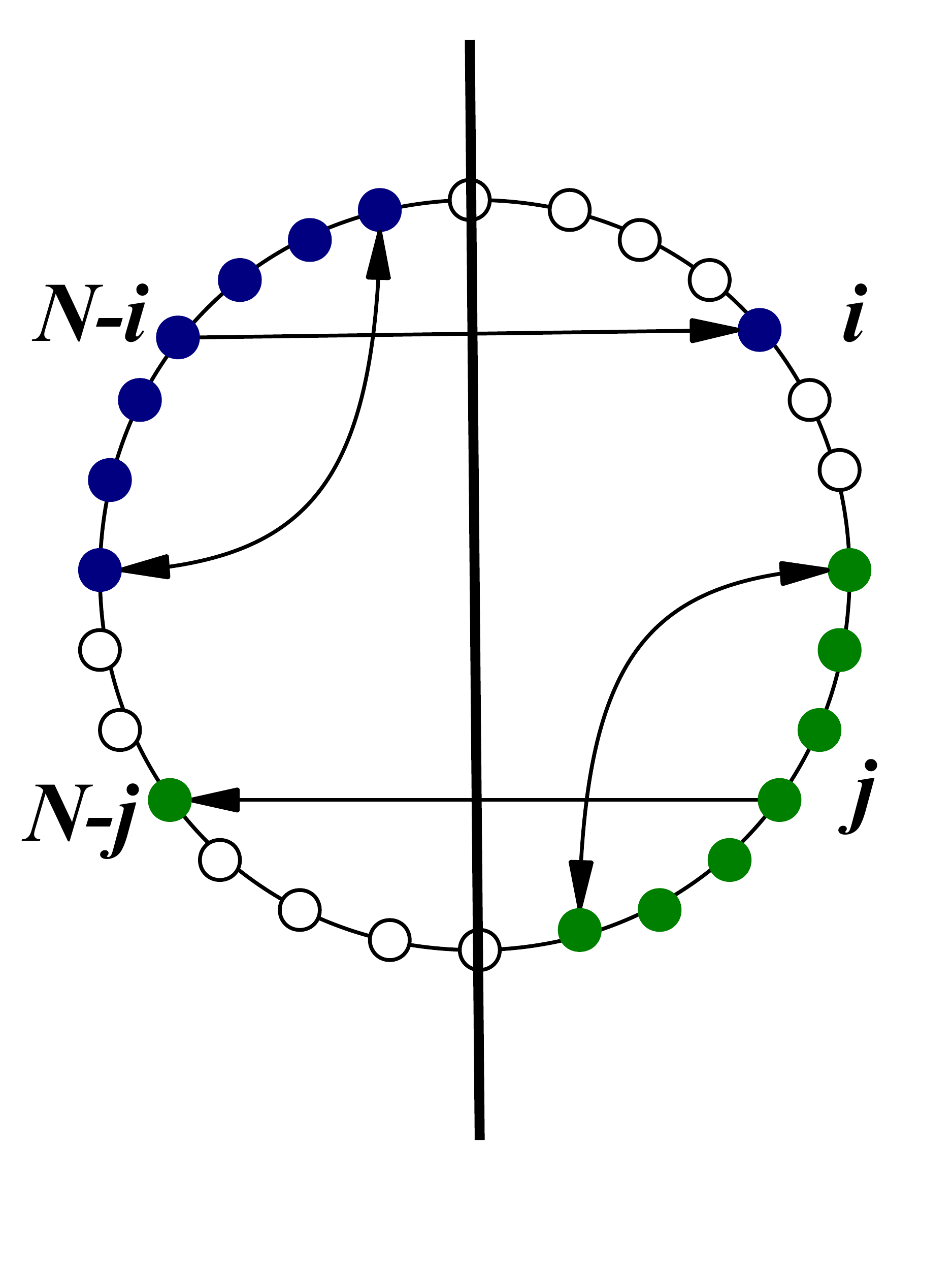}}
\caption{\label{figii-1} (Color online)
Schematic representation of the 
mirror connectivity around the diagonal passing from neurons set at positions N (top node)
 and $N/2$ (bottom node). For simplicity
only the connectivity of element at position $i$ on the top-right (blue elements) and of element $N-j$ on
the bottom-left (green) are depicted. The connectivity of all other elements (including the white ones)
 is not shown.
}
\end{figure}

In previous studies the influence of the complexity of the connectivity matrix on the steady state was
investigated. 
In refs.\cite{omelchenko:2015,hizanidis:2015,isele:2016,tsigkri:2015,ulonska:2016} hierarchical connectivity
was investigated giving rise to complex phenomena, such as hierarchical and traveling multichimeras and
transitions between chimera states with different multiplicity. In this study we introduce a new type of
connectivity, the mirror connectivity, or reflecting connectivity which is also inspired by the division
of the brain in two symmetric hemispheres. Mirror connectivity means that neurons belonging to one hemisphere
 can connect with neurons in the other hemisphere and vice-versa. In this simplistic model where the LIF
neurons are set in a ring architecture  we assume that
each neuron connects to its mirror image across a diagonal axis of the ring and to all the neurons belonging to
a region of size $R$ on the left and right of the mirror neuron, see Fig.~\ref{figii-1}.
Without loss of generality and because of the cyclic symmetry of our ring,
 we assume that the diagonal axis for the mirror images is the one
that passes from positions $k=N$ (top) and $k=N/2$ (bottom), as marked in Fig.~\ref{figii-1}. The reflecting
connectivity matrix
 $\sigma (i,j)$ linking node $j$ to $i$ takes the form:
\begin{eqnarray}
\sigma (i,j)=  
\left\{
  \begin{aligned}
    1 &\>\> {\rm for} \>\> (N-i-R)<j<(N-i+R)  \\
    0 &\>\> {\rm otherwise}
  \end{aligned}
\right.
\label{eq12}
\end{eqnarray}
\noindent Here also all indices are understood $\mod (N)$. 
The above rules allow each LIF element to interact with the same, constant number of $2R$ elements with directed
mirror links on the ring.

\begin{figure}[ht!]
\includegraphics[clip,width=0.97\linewidth,angle=0]{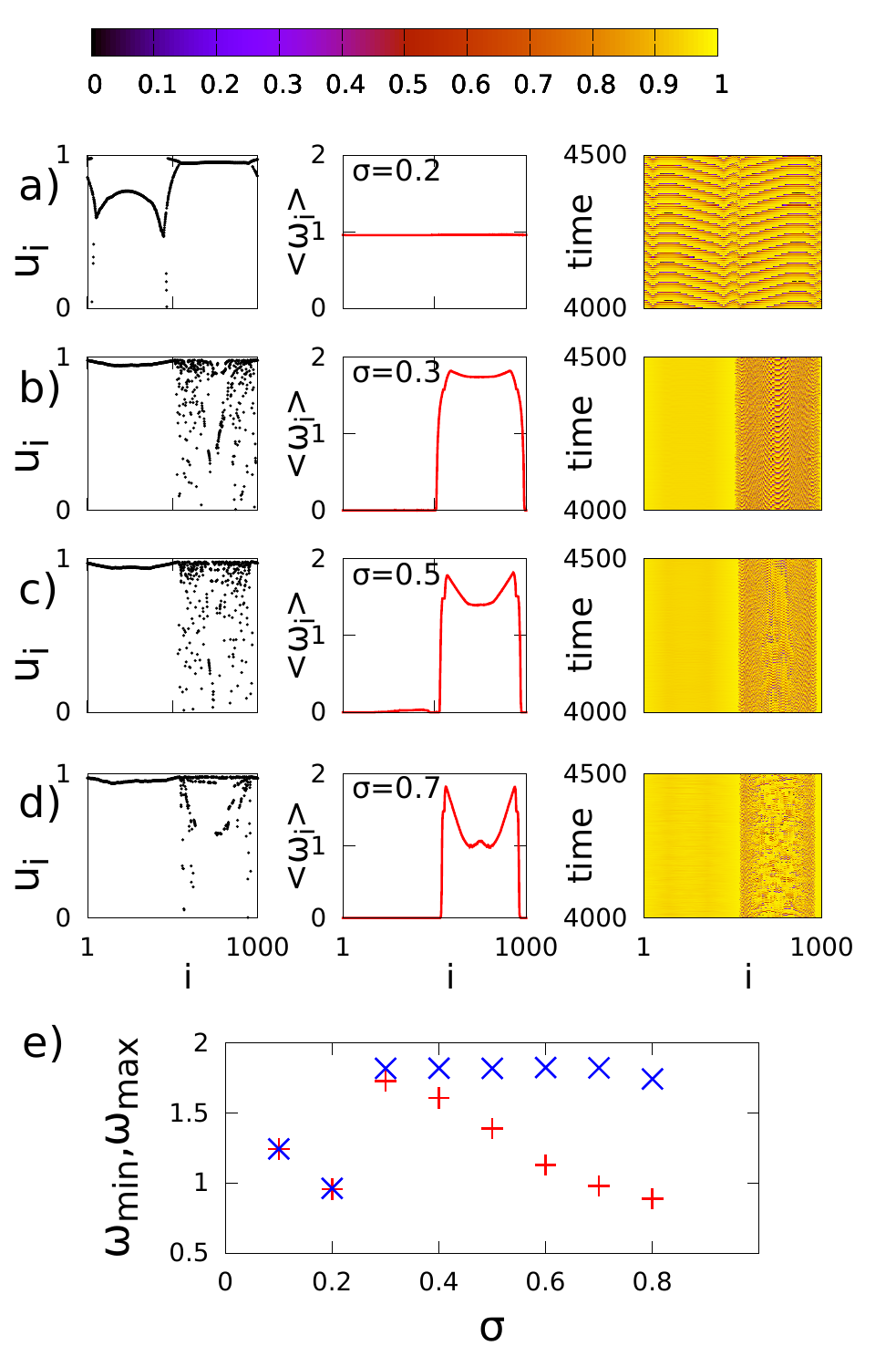}
\caption{\label{fig-ref2} (Color online)
LIF system with reflecting connectivity and varying coupling constant $\sigma$. Typical snapshots are
depicted in the left column, mean phase velocities in the middle and spacetime plots in the right column.
  a) $\sigma =0.2$, b) $\sigma =0.3$,  c) $\sigma =0.5$,  d) $\sigma =0.7$, 
e) $\omega_{min}$ and $\omega_{max}$ as a function of $\sigma$. 
Other parameters are $N=1000$, $R=100$, $\mu =1$ and $u_{\rm th}=0.98$ and $p_r=0$.
All realizations start from the same, randomly selected initial conditions between 0 and $u_{\rm th}$.
}
\end{figure}

 To investigate the effects of this connectivity scheme we fix the number of oscillators to $N=1000$ and the
number of links to $2R=200 \>(R=100)$ and 
the refractory period to $p_r$=0. The results are plotted in Fig.~\ref{fig-ref2}.
 Reflecting connectivity induces an unexpected phenomenon in the dynamics. 
From Fig.~\ref{fig-ref2} we see 
that for small values of the coupling constant $\sigma <0.3$ the system is synchronized and the
potential variables $u_i$ of all oscillators cover the entire range $0\le u_i < u_{\rm th}$, which is allowed from the system
constraints, Eq.~(\ref{eq01}). As $\sigma $ increases, we observe a transition around the value 0.3, where
the elements on the one half-ring remain close to the threshold while the 
ones in the other half-ring
oscillate with variable mean phase velocities. 
We can say that this is a spatial coexistence of a coherent steady state domain with incoherent oscillations,
 because on the one side the elements do not oscillate and on the other side they
oscillate but their mean phase velocities present modulations much like the ones observed in chimera
states. In fact, the $\omega$ variations are more complex than in classical chimeras;  as $\sigma$ 
increases away from the transition the oscillators in the middle of the active semi-ring develop gradually  
a convex shape with a minimum in the middle of the active semi-ring, Fig.~\ref{fig-ref2}c and d.
 The maximum value of the mean phase velocity remains constant, the depth of the convex shape increases
with $\sigma$ as can be seen in Fig.~\ref{fig-ref2}e, where the $\omega_{max}$ and $\omega_{min}$ values 
are depicted. As we increase further the coupling strength a secondary $\omega$ peak develops 
in the middle of the active region (Fig.~\ref{fig-ref2}d), where the oscillators speedup with respect
to their neighbors. For this value of coupling radius $R=100$ the size of the active region seems to
remain almost constant as a function of $\sigma$. This is not always the case,
 as we show evidence in the sequel.
 Incoherent states with complex modulation in their mean phase velocities were previously observed in the 
FitzHugh-Nagumo model, when the single chimeras are transformed into multichimeras as the 
coupling range $r$ increases \cite{omelchenko:2013}. The difference with the current study
 is that here the multichimera
states coexist with a number of near-threshold oscillators and this is due to the combined effect
of reflecting connectivity and positive (attracting) coupling dynamics. Another difference to Ref.~\cite{omelchenko:2013} is that the modulations here take place when the coupling constant $\sigma$ changes, while in Ref.~\cite{omelchenko:2013} multiplicity of the chimera increases as $r$ increases.

\begin{figure}[ht!]
\includegraphics[clip,width=0.97\linewidth,angle=0]{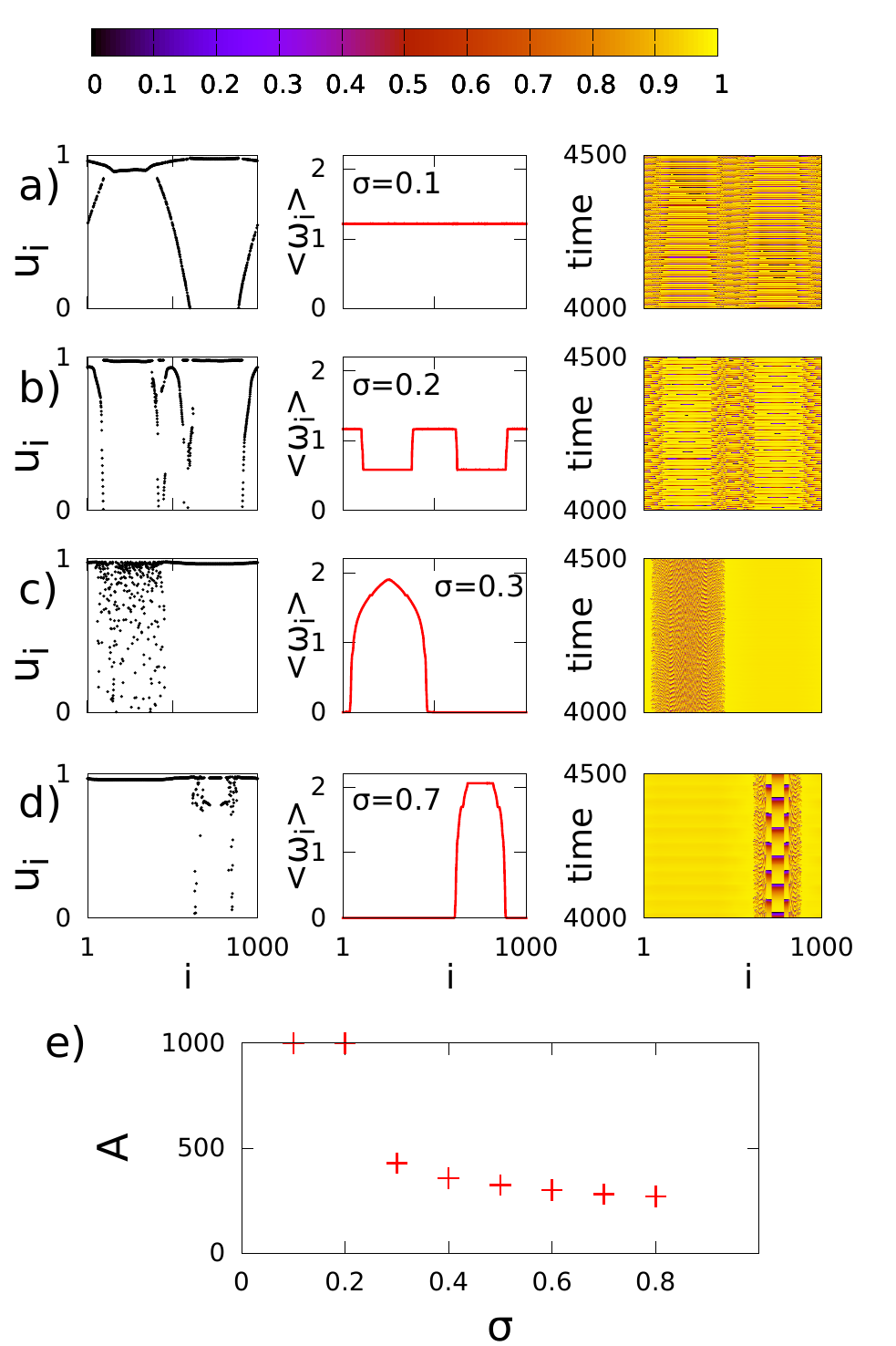}
\caption{\label{fig-ref4} (Color online)
LIF system with reflecting connectivity similar to Fig.~\ref{fig-ref2} and $R=300$.
  a) $\sigma =0.2$, b) $\sigma =0.3$,  c) $\sigma =0.5$,  d) $\sigma =0.7$, 
e) Size of the active region $A$ as a function of $\sigma$.
All other parameters as in Fig.~\ref{fig-ref2}.
}
\end{figure}

 To explore further the influence of this connectivity in the system synchronization, 
we choose to examine the same features for a large value of coupling radius, $R=300$. Figure \ref{fig-ref4} 
shows that for small values of
$\sigma$ we have the synchronized regime, around $\sigma =0.2$ a first destabilization takes place
where the system splits into two pairs of mutually synchronous regions, around $\sigma =0.3$ only one
of the semi-rings stays active, while in the other one the oscillators stay near the threshold and do 
not venture away from it. As before, near the threshold the mean phase velocity profile on the semi-ring achieves
an arc-like shape, while as $\sigma$ increases a plateau starts developing. For even larger values of $\sigma$
the convex shape is formed in the middle of the active region (not shown). Comparison between panels 
 \ref{fig-ref4}c and \ref{fig-ref4}d indicates that the active region, 
where excursions away from the threshold take place, 
shrinks with $\sigma$. Figure \ref{fig-ref4}e depicts the size of the active region $A$ as a 
function of $\sigma$. For small $\sigma$ all $N=1000$ elements are active. Around $\sigma =0.3$ a transition
takes place and the active region is confined in one of the two semi-rings. The active region shrinks 
further as $\sigma$ increases.

\begin{figure}[ht!]
\includegraphics[clip,width=0.97\linewidth,angle=0]{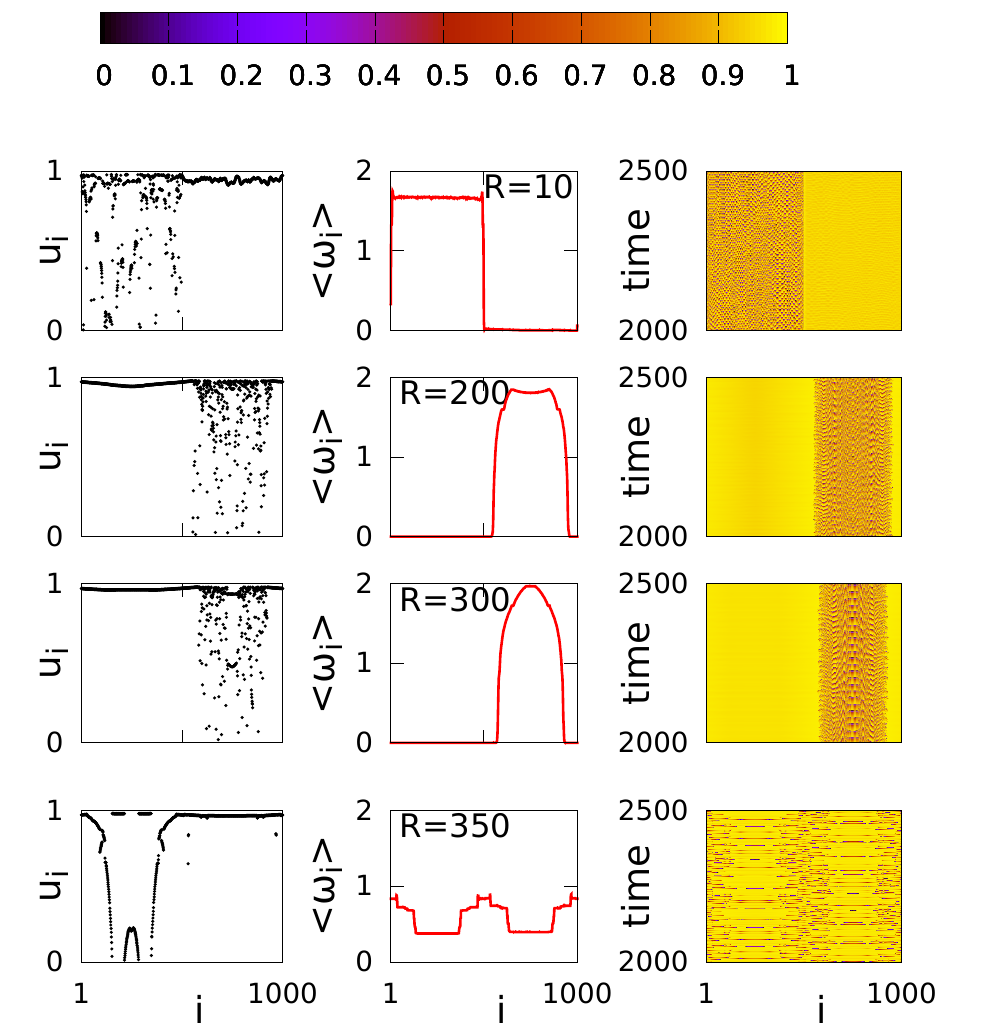}
\caption{\label{fig-ref7} (Color online)
LIF system with reflecting connectivity with varying coupling radius $R$. Typical snapshots are
depicted in the left column, mean phase velocities in the middle and spacetime plots in the right column.
  a) $R =10$, b) $R=200$,  c)  $R=300 $, d) $R=350 $ and e) $R =400$. $\sigma =0.4$ $p_r=0$ and other
parameters as in Fig.  \ref{fig-ref2}.
}
\end{figure}

 To prompt further into the properties of this system we study
 the dependence on the coupling radius $R$, keeping the coupling strength constant, $\sigma =0.4$, and other parameter values as in the previous figures. In Fig.~\ref{fig-ref7}
we can see that even for small $R$-values, as small as 10, the activity is isolated into one of the two
semi-rings, with abrupt change of behaviour at the interfaces.
Counter-intuitively, as the coupling radius increases the width of the active stripe decreases up to a limit value of $R$, which for these parameter values is in the range $300 \le R_{\rm lim} \le 350$, 
Fig.~\ref{fig-ref7}c. 
Note that for this limiting value the $\omega$-shape in the active semi-ring takes
the arc form. 
For large values of $R>R_{\rm lim}$ the spreading of the coupling in larger regions causes oscillations
in both parts of the system, Fig.~\ref{fig-ref7}. The symmetry in the reflecting connectivity can be seen in the mean phase
velocity plots: the oscillations in both semi-rings are images of one another. Coherent oscillations
are developed in the middle of each semi-ring while the elements which border the semi-ring connections
are out-of-order (incoherent). As we increase further the coupling radius the incoherent regions shrink,
while the system synchronizes completely.
In this respect,  increasing the coupling radius or the coupling constant causes different
effects in the system: when the coupling radius $R$ increases the system passes from (active-passive) semi-rings
 to (active-active) semi-rings; when the coupling constant $\sigma $ increases the system passes from
the (active-active) phase to the (active-passive) one with simultaneous shrinking of the active phase.

\begin{figure}[ht!]
\includegraphics[clip,width=0.97\linewidth,angle=0]{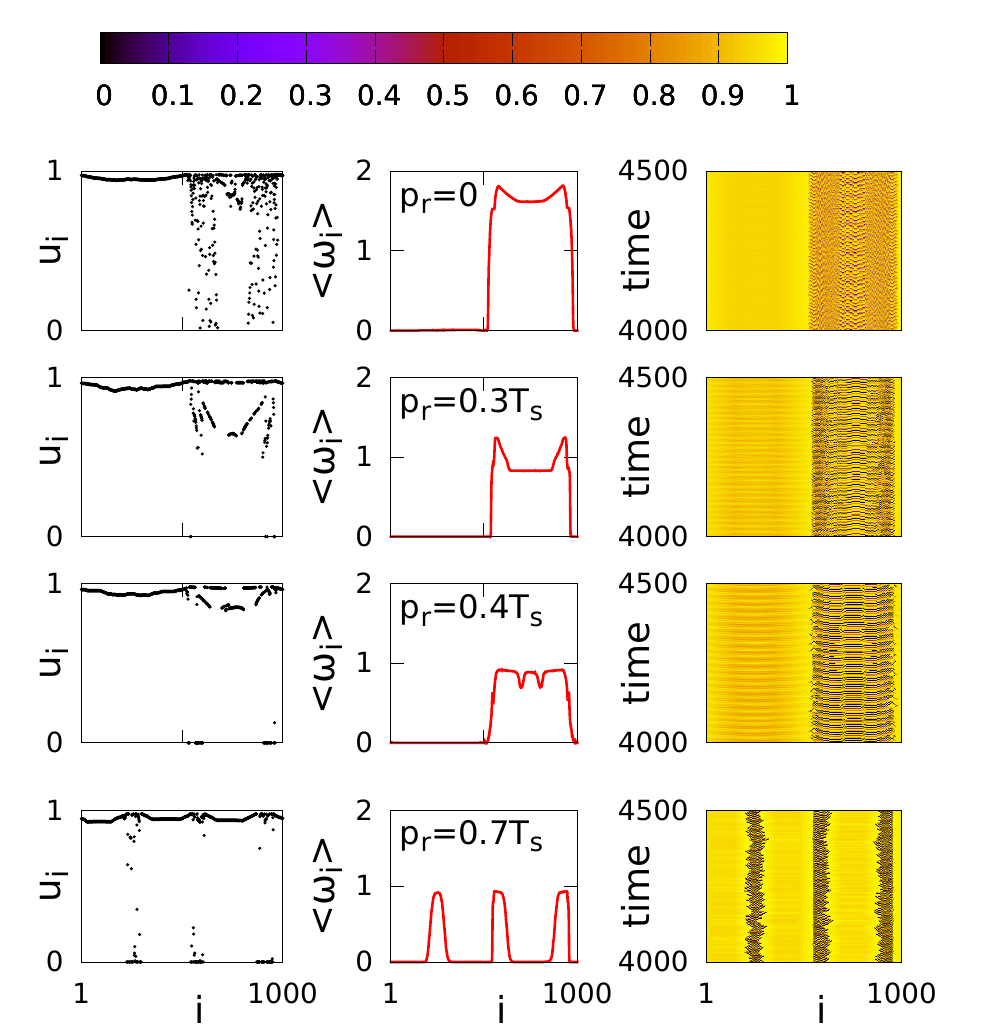}
\caption{\label{fig-ref5} (Color online)
LIF system with reflecting connectivity with varying refractory period $p_r$. Typical snapshots are
depicted in the left column, mean phase velocities in the middle and spacetime plots in the right column.
  a) $p_r =0$, b) $p_r = 0.3 T_s $,  c) $p_r =0.4 T_s $, 
and  d) $p_r =0.7T_s $. $R=100$, $\sigma =0.4$ and other
parameters as in Fig.  \ref{fig-ref2}.
}
\end{figure}

 We next consider variations in the refractory period $p_r$ in Fig.~\ref{fig-ref5} keeping other parameters fixed to $R=100$ and $\sigma =0.4$.
For $p_r=0$  the system keeps the usual convex semi-ring arrangement as was also observed previously for the
variations with $R$, see Fig.~\ref{fig-ref7}. As $p_r$ increases the convex arrangement deepens, while around
$p_r \sim 0.4T_s$  the mean phase velocity develops 2 minima of similar structure. This behaviour persists
for larger values of $p_r$, until around $p_r\sim 0.7T_s$ where a second transition takes place and 
the active domain splits into three smaller active ones which now reside in both semi-rings.
The introduction of the refractory period thus
 induces multiple coherent and incoherent regions, separated by the
near-threshold elements. These three regions persist for even larger values of $p_r$. Transient
states (not shown) with two active stripes were observed 
but they were short-living turning into either the
single stripe or to the
structure with three stripes.

\begin{figure}[ht!]
\includegraphics[clip,width=0.97\linewidth,angle=0]{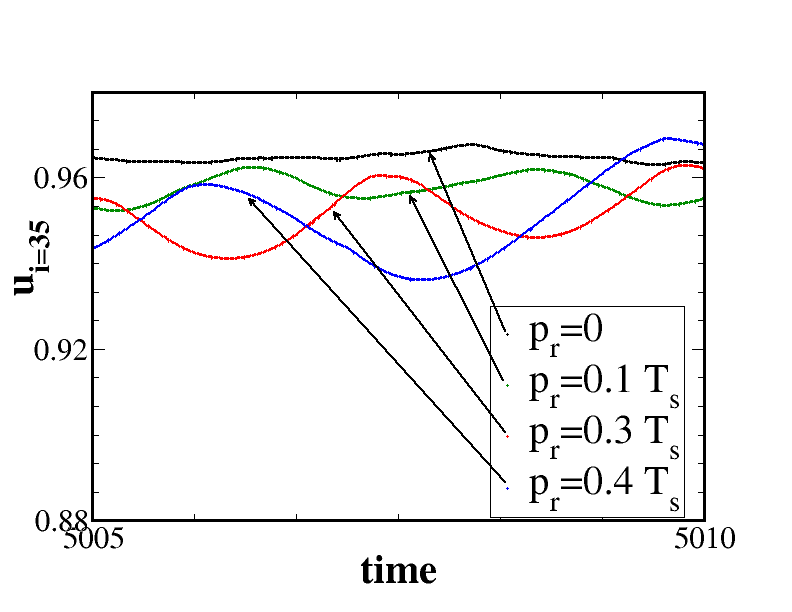}
\caption{\label{fig-ref6} (Color online)
Temporal evolution of the potential $u$ of threshold LIF elements. Lines depict the fluctuations of 
element $i=35$, for $p_r=0$ (black line), $p_r=0.1 T_s$ (green line)
 $p_r=0.3 T_s$ (red line) and $p_r=0.4 T_s$ (blue line). $R=100$, $\sigma =0.4$ and other
parameters as in Fig.  \ref{fig-ref5}.
}
\end{figure}

 Finally, we take a closer look at the near-threshold elements. These elements are not entirely fixed
 but fluctuate near
the threshold potential. Fig.~\ref{fig-ref6} shows the temporal behaviour of the element $i=35$ in the cases
of $p_r=0, \> 0.1T_s, \> 0.3T_s$ and $0.4 T_s$, with all other parameters as in Fig.~\ref{fig-ref5}. 
This figure provides evidence that the threshold elements fluctuate with higher 
amplitude as the refractory period increases.  

 Overall, the introduction of a reflecting connectivity architecture induces restriction of the oscillations on a single domain which resides in one half of the ring, 
while in the other half the elements remain close to the threshold potential $u_{\rm th}$. 
Within the active domains the oscillators organize in domains with different phase
velocities and transition regions. 
The introduction of a common refractory period to all LIF elements induces 
splitting of the single active domain into smaller ones; 
these smaller active stripes reside in both ring semi-circles.  

\section{Conclusions}
\label{conclusions}
Synchronisation phenomena in diffusively coupled Leaky Integrate-and-Fire elements are
 examined in the case of
nonlocal coupling and reflecting connectivity. In the case of the classical nonlocal 
diffusive coupling traveling
waves of synchronous elements with constant phase difference are observed, while chimera
states are difficult to identify.
Due to the travelling of the waves all elements spent some part of the time in the coherent and part
in the incoherent regions and thus the difference in the mean phase velocity, which is 
the signature of the chimera states,  smears out. The number of oscillators which are below the
threshold is statistically constant in time and only depends on the parameter values $\sigma$ and $p_r$. 
In the case of reflecting connectivity
evidence is produced for novel incoherent domains coexisting with near threshold elements.
This phenomenon is observed in the coupled  model when each element is connected
with elements in its mirror across a diagonal axis. For this connectivity, the oscillations are restricted in
one half of the ring, while the elements of the other half stay near the threshold. For small values
of the  coupling constant $\sigma$ the oscillatory activity covers the entire ring. As $\sigma$ 
increases the activity is restricted in one of the semi-rings while the size of the active region 
decreases inversely with $\sigma$. If a refractory period is added to the system
the single activity regions split, the active regions are spread in both semi-rings and they are
separated by near-threshold elements
\par
  In the current study we assume that all elements are identical, i.e., they have common parameters: 
the coupling constant $\sigma$, the refractory
period $p_r$, the threshold potential $u_{\rm th}$ and the accumulating rate $\mu$. In further studies
these assumption can be relaxed and different parameters can be assigned to the elements/neurons, 
depending on their physiology and
functioning. Furthermore, one can add inhomogeneity in the form of a small dispersion around a
mean value, to any of these parameters and study the implications of this statistical variation
 in the form and
properties of the chimera states.
\par
 Topology-connectivity is also relevant in neurological disorders (schizophrenia, Alzheimer, Parkinson
etc). In some of them the damage may be hereditary, in others it may develop with age. 
In all these cases
it is important to understand how connectivity changes with previous experience, disorder and age and how this influences/modifies the network synchronization patterns. 
The outcome of research in this direction 
will be twofold: on the one hand the relation between neuronal network architecture 
and brain functionality (perception and memory) can be established and on the other hand
the association of connectivity with brain disorders will be elucidated. 
MRI and resting vs. task oriented fMRI experiments are expected to shed light in these important issues.

\section{Acknowledgments}
This work was supported by the German Academic Exchange Service (DAAD) 
and the Greek State Scholarship Foundation IKY within the PPP-IKYDA framework.
Funding was also provided by NINDS R01-40596. The research work was partially
supported by the European Union's Seventh Framework Program (FP7-REGPOT-2012-2013-1) 
under grant agreement n316165. 
PH and ES
acknowledge support by DFG in the framework of the Collaborative Research Center 910. 
JH acknowledge support in the framework of the SIEMENS program ``Establishing a Multidisciplinary 
and Effective Innovation and Entrepreneurship Hub''.
This work was 
supported by computational time granted from the Greek Research \& Technology Network (GRNET)
in the National HPC facility - ARIS - under project ID PA002002.

\end{document}